\documentclass[12pt]{iopart}

\newcommand{\be}{\begin{equation}}
\newcommand{\ee}{\end{equation}}
\newcommand{\bea}{\begin{eqnarray}}
\newcommand{\eea}{\end{eqnarray}}
\newcommand{\beaa}{\begin{eqnarray*}}
\newcommand{\eeaa}{\end{eqnarray*}}
\newcommand{\nn}{\nonumber \\}

\begin{document}

\tolerance=5000

\title{Dark energy problem: from phantom theory to modified Gauss-Bonnet gravity}

\author{Shin'ichi Nojiri
\footnote{Electronic address: snojiri@yukawa.kyoto-u.ac.jp,
nojiri@cc.nda.ac.jp}}
\address{Department of Applied Physics, National Defence Academy, Hashirimizu Yokosuka 239-8686, Japan}

\author{Sergei D. Odintsov
\footnote{Electronic address: odintsov@ieec.uab.es,
also at TSPU, Tomsk}}
\address{Instituci\`{o} Catalana de Recerca i Estudis Avan\c{c}ats (ICREA)
and Institut d' Estudis Espacials de Catalunya (IEEC/ICE),
Edifici Nexus, Gran Capit\`{a} 2-4, 08034 Barcelona, Spain}

\author{O.G. Gorbunova}
\address{Lab. for Fundamental Studies,
Tomsk State Pedagogical University,
634041 Tomsk, Russia}

\begin{abstract}

The solution of the dark energy problem in  models without scalars is
presented. It is shown that a late-time accelerating cosmology may be
generated by an ideal fluid with some implicit equation of state.
The  evolution of the universe within modified Gauss-Bonnet gravity is 
considered.
It is demonstrated that such gravitational approach may predict the
(quintessential, cosmological constant or transient phantom) acceleration
of the late-time
universe with a natural transition from deceleration to acceleration (
or from non-phantom to phantom era in the last case).

\end{abstract}

\maketitle

\section{Introduction}

According to recent astrophysical data the current universe
is expanding with acceleration. Such accelerated behaviour of the universe
is supposed to be due to the presence of mysterious dark energy which
,at present, contributes about 70\% of the total universe energy-mass. 
What  this dark
energy is and where it came from is one of the fundamental problems
of modern theoretical cosmology (for a recent review, see
\cite{nabhan}). Assuming a constant
equation of state (EOS),
$p=w\rho$, the dark energy may be associated with a (so far unobserved ) 
strange
ideal fluid with negative w. Astrophysical data indicate that w lies in
a very narrow strip close to $w=-1$.
The case $w=-1$  corresponds to the cosmological constant.
For $w$ less than $-1$ the phantom dark energy is observed, and for 
$w$ more than
$-1$ (but less than $-1/3$)  the dark energy is described by quintessence.
It is interesting that the phantom phase is twice as probable than
the quintessence phase. Moreover, there are indications that there occured
a recent transition over cosmological constant barrier (over the phantom
divide). In this case, which is not completely confirmed yet, the universe 
lives
in its phantom phase which ends eventually
at a future singularity (Big Rip) \cite{mcinnes}.

There are various approaches to describe the phantom dark energy.
The simplest one is to work with negative kinetic energy scalar (phantom).
There may be several scalars in the theory, with at least one of them 
being
phantom. The presence of a scalar potential or a non-minimal coupling with
gravity is also necessary.
The scalar phantom cosmology has been studied in a number of works
(for recent discussion see \cite{phantom,phantom2,phantom3,QG}).
The typical property of this phantom cosmology is a future singularity
which occurs in a finite time. This is due to the growth of phantom energy
which may lead to quite spectacular consequences. In particular,
with growth of phantom energy,  typical energies (as well as
curvature invariants) increase in the expanding universe. As a result, in some scenarios
the quantum gravity era may come back at the end of the phantom
universe evolution. In this case, it was checked that quantum
effects\cite{quantum,QG,tsujikawa} may act against the Big Rip (or even stop it
in case of quantum gravity\cite{QG}).
In principle, there are four types of future singularities which are
classified in  \cite{tsujikawa}.

The emerging phantom era may exist in string-inspired gravity where
typically a scalar field as well as a Gauss-Bonnet term coupled to a 
scalar
are included \cite{stringy}. String considerations may also lead to
effective (phantom-like) theories \cite{stringy1}. Finally, gravitational
theory with a time-dependent cosmological constant\cite{lambda} may also
mimic the phantom regime.

The interesting approach to describe the dark energy universe is
related to an ideal fluid with some (strange but 
explicit)
EOS which may be sufficiently
complicated \cite{EOS}. Of course, this approach is phenomenological
in some sense, because it does not describe the fundamental origin
of dark energy. At the same time, it may lead to quite successful
description of not only phantom phase but also of transition from
decelation to acceleration or crossing of the phantom divide.
Moreover, there is no need to introduce  scalars in cosmology.
One can generalize it assuming inhomogeneous EOS of the universe \cite{inh}
which may be interpreted also as some modification of GR.
In the present work we discuss two dark energy models ( one with
implicit EOS of the universe and  one where gravity is modified by
the function of Gauss-Bonnet (GB) term).
It is shown that such models lead to successful (quintessence,
cosmological constant or phantom)
late-time acceleration.

\section{The acceleration due to dark energy with implicit EOS}

We now consider the FRW cosmology with an ideal fluid.
The starting FRW universe metric is:
\be
\label{FRW}
ds^2=-dt^2 + a(t)^2 \sum_{i=1}^3  \left(dx^i\right)^2\ .
\ee
In the FRW universe, the energy conservation law can be expressed as
\be
\label{ppH1}
0=\dot\rho + 3H\left(p + \rho\right)\ .
\ee
Here $\rho$ is the energy density,and $p$ is the pressure. The Hubble rate 
$H$ is defined
by $H\equiv \dot a/a$ and
  the first FRW equation is
\be
\label{pH3}
\frac{3}{\kappa^2}H^2=\rho\ .
\ee
We often consider the case that $\rho$ and $p$ satisfy the simple EOS, $p=w\rho$.
Then if $w$ is a constant, Eq.(\ref{ppH1}) can be easily integrated as
$\rho=\rho_0 a^{-3(1+w)}$.
Using the first FRW equation (\ref{pH3}),  the well-known solution
follows:
\be
\label{ppH4}
a=a_0 \left(t - t_1\right)^\frac{2}{3(w+1)}\quad \mbox{or}
\quad a_0 \left(t_2 - t\right)^\frac{2}{3(w+1)}\ ,
\ee
when $w\neq -1$, and
\be
\label{ppH5}
a=a_0\e^{\kappa t\sqrt{\frac{\rho_0}{3}}}
\ee
when $w=-1$.

The ideal fluid with a more general EOS may be considered\cite{EOS}:
\be
\label{EoS1}
f(\rho, p)=0\ .
\ee
An interesting example is given by
\be
\label{EoS2}
\rho\left(1 + \frac{A}{2}(\rho + p)\right)
+ \frac{3}{\rho}\left(2 + \frac{A}{2}(\rho + p)\right)^2 =0\ .
\ee
Solving Eqs.(\ref{ppH1}) and  (\ref{pH3}) one arrives at
\bea
\label{EoS3}
&& H=\frac{1}{t}\left( 1 - \frac{\kappa^2t^2}{A}\right)\ ,\quad
\rho = \frac{3}{\kappa^2t^2}\left( 1 - \frac{\kappa^2t^2}{A}\right)^2\ ,\nn
&& p=\frac{1}{\kappa^2}\left(\frac{1}{t^2} - \frac{8\kappa^2}{A} + \frac{3\kappa^4 t^2}{A^4}\right)\ .
\eea
Since
\be
\label{EoS4}
\dot H= - \frac{1}{t^2} - \frac{\kappa^2}{A}\ ,
\ee
it follows that
\be
\label{EoS5}
\frac{\ddot a}{a}=\dot H + H^2 = \frac{\kappa^2}{A}\left(- 3 + \frac{\kappa^2 t^2}{A}\right)\ .
\ee
Then if $A>0$, the decelerating universe with $\ddot a<0$ transits to an 
accelerating universe
with $\ddot a>0$ when $t=\sqrt{\frac{3A}{\kappa^2}}$.
Eq.(\ref{EoS4}) also shows that if $A<0$, the non-phantom phase with $\dot
H<0$ changes to the
phantom phase with $\dot H>0$ at $t=\sqrt{-\frac{A}{\kappa^2}}$.
This demonstrates that an ideal fluid with a complicated EOS may be the 
origin
of dark energy
and late-time acceleration.

\section{Dark energy from modified Gauss-Bonnet gravity}

In this section, we consider the following gravitational action\cite{GB}:
\be
\label{GB1}
S=\int d^4 x\left(\frac{1}{2\kappa^2}R + f(G) + {\cal L}_m\right)\ .
\ee
Here ${\cal L}_m$ is the matter Lagrangian density and
$G$ is the GB invariant:
$G=R^2 -4 R_{\mu\nu} R^{\mu\nu} + R_{\mu\nu\xi\sigma} R^{\mu\nu\xi\sigma}$.
Such a theory has a nice Newtonian limit and may be considered as 
alternative
for GR \cite{GB}.
The variation over $g_{\mu\nu}$ gives:
\bea
\label{GB4b}
&& 0= \frac{1}{2\kappa^2}\left(- R^{\mu\nu} + \frac{1}{2} g^{\mu\nu} R\right)
+ T^{\mu\nu} + \frac{1}{2}g^{\mu\nu} f(G) -2 f'(G) R R^{\mu\nu} \nn
&& + 4f'(G)R^\mu_{\ \rho} R^{\nu\rho} - 2 f'(G) R^{\mu\rho\sigma\tau}R^\nu_{\ \rho\sigma\tau}
  -4 f'(G) R^{\mu\rho\sigma\nu}R_{\rho\sigma} \nn
&& + 2 \left( \nabla^\mu \nabla^\nu f'(G)\right)R
  - 2 g^{\mu\nu} \left( \nabla^2 f'(G)\right)R
  - 4 \left( \nabla_\rho \nabla^\mu f'(G)\right)R^{\nu\rho} \nn
&& - 4 \left( \nabla_\rho \nabla^\nu f'(G)\right)R^{\mu\rho}
+ 4 \left( \nabla^2 f'(G) \right)R^{\mu\nu} \nn
&& + 4g^{\mu\nu} \left( \nabla_{\rho} \nabla_\sigma f'(G) \right) R^{\rho\sigma}
+ 4 \left(\nabla_\rho \nabla_\sigma f'(G) \right) R^{\mu\rho\nu\sigma} \ .
\eea
The analog of the first FRW equation has the following form:
\be
\label{GB7}
0=-\frac{3}{\kappa^2}H^2 + Gf'(G) - f(G) - 24 \dot Gf''(G) H^3 + \rho_m\ .
\ee
Here $\rho_m$ is the matter energy density.
%For (\ref{FRW}), GB invariant $G$ has the following form:
%\be
%\label{GB8}
%G=24\left(\dot H H^2 + H^4\right)\ .
%\ee
When $\rho_m=0$, Eq.(\ref{GB7}) has a deSitter universe solution where $H$ 
and
therefore $G$ are constants.
If $H=H_0$ with constant $H_0$, Eq.(\ref{GB7}) looks as\cite{GB}:
\be
\label{GB7b}
0=-\frac{3}{\kappa^2}H_0^2 + 24H_0^4 f'\left( 24H_0^4 \right) - f\left( 24H_0^4\right) \ .
\ee
For a large number of choices of the function $f(G)$, Eq.(\ref{GB7b}) has 
a
non-trivial ($H_0\neq 0$) real solution for $H_0$ (deSitter universe).

We now consider the case $\rho_m\neq 0$.
Let the EOS parameter $w\equiv p_m/\rho_m$ be a constant.
Using the conservation of  energy: $\dot \rho_m + 3H\left(\rho_m +
p_m\right)=0$,
one finds $\rho=\rho_0 a^{-3(1+w)}$.
Assume $f(G)$ is given by
\be
\label{mGB1}
f(G)=f_0\left|G\right|^\beta\ ,
\ee
with constants $f_0$ and $\beta$.
If $\beta<1/2$, $f(G)$ term becomes dominant compared with the Einstein term when the
curvature is small. In this case, the solution is
\be
\label{mGB2}
a=\left\{\begin{array}{ll} a_0t^{h_0}\ &\mbox{when}\ h_0>0 \\
a_0\left(t_s - t\right)^{h_0}\ &\mbox{when}\ h_0<0 \\
\end{array} \right. \ ,
\ee
where
\bea
\label{mGB3}
h_0&=&\frac{4\beta}{3(1+w)}\ ,\nn
a_0&=&\Bigl[ -\frac{f_0(\beta - 1)}{\left(h_0 - 1\right)\rho_0}\left\{24 \left|h_0^3 \left(- 1 + h_0\right)
\right|\right\}^\beta
\left( h_0 - 1 + 4\beta\right)\Bigr]^{-\frac{1}{3(1+w)}}\ .
\eea
One may define the EOS parameter $w_{\rm eff}$
by
\be
\label{FRW3k}
w_{\rm eff}=\frac{p}{\rho}= -1 - \frac{2\dot H}{3H^2}\ ,
\ee
which is less than $-1$ if $\beta<0$ and $w>-1$ as
\be
\label{mGB3b}
w_{\rm eff}=-1 + \frac{2}{3h_0}=-1 + \frac{1+w}{2\beta}\ .
\ee
If $\beta<0$, we obtain the effective phantom phase with negative $h_0$
even if
$w>-1$.
In the phantom phase,  the Big Rip type singularity at $t=t_s$ might 
occur.
Near the Big Rip singularity, however, the curvature becomes dominant
  and $f(G)$-term may be neglected. Then the universe expands as
$a=a_0t^{2/3(w+1)}$.
Hence, the Big Rip singularity  eventually does not occur.

A similar model has been constructed in \cite{ANO} within a consistent
modified
$f(R)$-gravity\cite{modified}.
In the case of $f(R)$-gravity,  instabilities appear in general 
\cite{DK}.
Such instabilities are not common in $f(G)$-gravity.

We should note that under the assumption (\ref{mGB2}), the GB invariant $G$ and scalar curvature $R$
behave as
\bea
\label{mGB4}
G&=&\frac{24h_0^3 \left(h_0 - 1\right)}{t^4}\ \mbox{or}\
\frac{24h_0^3 \left(h_0 - 1\right)}{\left(t_s -t\right)^4}\ , \nn
R&=&\frac{6h_0 \left(2h_0 - 1\right)}{t^2}\ \mbox{or}\
\frac{6h_0 \left(2h_0 - 1\right)}{\left(t_s -t\right)^2}\ .
\eea
When the scalar curvature $R$ is small, that is,
$t$ or $t_s - t$ is large, the GB invariant $G$ becomes smaller more
rapidly
than $R$. When $R$ is large, that is, $t$ or $t_s - t$ is small,
$G$ becomes larger more rapidly than $R$. Hence, if $f(G)$ is given by
(\ref{mGB1}) with $\beta<1/2$,
the $f(G)$-term in the action (\ref{GB1}) becomes  dominant for
  small curvature
  but becomes less dominant for  large curvature.
Eq.(\ref{mGB3}) follows when the curvature is small.
There are, however, some exceptions. As clear from  (\ref{mGB4}),
when $h_0=-1/2$, which corresponds to $w_{\rm eff}=-7/3$, $R$ vanishes and
when $h_0=-1$, corresponding to $w_{\rm eff}=-5/3$, $G$ vanishes. In these cases,
only one of the Einstein and $f(G)$ terms survives.

One may also consider the case that $0<\beta<1/2$. As $\beta$ is positive,
the universe does not reach
the phantom phase. When the curvature is strong, the $f(G)$-term
  could be neglected.
  If $w$ is positive, the matter energy density $\rho_m$
should behave as $\rho_m\sim t^{-2}$ but $f(G)$ behaves
  as $f(G)\sim t^{-4\beta}$. Thus, at a late time
(large $t$), the $f(G)$-term dominates. Without the contribution
from matter, Eq.(\ref{GB7}) has a deSitter universe solution where $H$ and 
therefore $G$ are constants.
If $H=H_0$ with constant $H_0$, Eq.(\ref{GB7}) looks as (\ref{GB7b}).
Hence, even if one starts from the deceleration phase with $w>-1/3$, we
may reach the asymptotically
deSitter universe, which is accelerated universe. In other words, there
occurs a transition from
  deceleration to acceleration in the universe evolution.

Now we consider the case where the contributions from the Einstein term 
and matter may be neglected.
  Eq.(\ref{GB7}) reduces to
\be
\label{mGB9}
0=Gf'(G) - f(G) - 24 \dot Gf''(G) H^3 \ .
\ee
If $f(G)$ behaves as (\ref{mGB1}), by the assumption (\ref{mGB2}), it
follows
\be
\label{mGB10}
0=\left(\beta - 1\right)h_0^6\left(h_0 - 1\right) \left(h_0 - 1 + 4\beta \right)\ .
\ee
As $h_0=1$ implies $G=0$, we may choose $h_0 = 1 - 4\beta$.
Eq.(\ref{FRW3k}) gives $w_{\rm eff}=-1 + 2/3(1-4\beta)$.
Therefore if $\beta>0$, the universe is accelerating ($w_{\rm eff}<-1/3$) and
if $\beta>1/4$, the universe is in phantom phase ($w_{\rm eff}<-1$).
The following model may be also constructed:
\be
\label{mGB13}
f(G)=f_i\left|G\right|^{\beta_i} + f_l\left|G\right|^{\beta_l} \ .
\ee
Here it is assumed
\be
\label{mGB14}
\beta_i>\frac{1}{2}\ ,\quad \frac{1}{2}>\beta_l>\frac{1}{4}\ .
\ee
When the curvature is large, as in the primordial universe, the first term
dominates and
\be
\label{mGB15}
-1>w_{\rm eff}=-1 + \frac{2}{3(1-4\beta_i)}>-5/3\ .
\ee
On the other hand, when the curvature is small, as in the present universe, the second term
in (\ref{mGB13}) dominates  and gives
\be
\label{mGB16}
w_{\rm eff}=-1 + \frac{2}{3(1-4\beta_l)}<-5/3\ .
\ee
Thus the modified GB gravity  (\ref{mGB13}) may describe the unification
of the
inflation and the late-time
acceleration of the universe (compare with proposal\cite{CNO}).

Instead of (\ref{mGB14}), one may choose $\beta_l$ as $\frac{1}{4}>\beta_l>0$, which gives
$-\frac{1}{3}>w_{\rm eff}>-1$.
Then the effective quintessence follows. By properly adjusting the
couplings $f_i$ and $f_l$
in (\ref{mGB13}), one may obtain a period where the Einstein term
dominates and the universe is in a 
deceleration phase. After that, a transtion from
deceleration to acceleration occurs
when the GB term  dominates.

Let us consider the system (\ref{mGB13}) coupled with matter as in
(\ref{GB7}).
We now choose $\beta_i>\frac{1}{2}>\beta_l$.
When the curvature is large, as in the primordial universe, the first term
dominates compared
with the second and the Einstein terms. When the curvature is small, as in
the present universe,
the second term in (\ref{mGB13}) dominates compared with the second and the Einstein terms.
The effective $w_{\rm eff}$ is given by (\ref{mGB3b}). In the primordal
universe,
the matter was radiation with $w=1/3$, and then the effective $w$ is given 
by
$w_{i,{\rm eff}}= -1 + 2/3\beta_i$,
which could be less than $-1/3$. That is, the universe is accelerating 
when $\beta_i>1$.
On the other hand, in the late-time universe matter could be dust with 
$w=0$, and
$w_{l,{\rm eff}}=-1 + \frac{1}{2\beta_l}$,
which is larger than $0$ if $0<\beta_l<1/2$ or less than $-1$ if $\beta_l$
is negative. Hence, in order that the acceleration of the universe could 
occur
in both the primordial and late-time universe, the conditions
$\beta_i>1$ and $\beta_l<0$ are required.
In the same way, other specific models of a late-time accelerating 
universe
may be constructed in frames of modified GB gravity without use of scalar
fields.

The modified GB gravity cosmology remains to be studied in more detail,
comparing it with observational data (in the same style as
in\cite{multamaki}) , investigating the perturbations
structure in the analogy with such study in other modified gravities
\cite{koivisto}. One has to compare carefully the predictions of modified
GB gravity with solar system tests in order to find the conditions to the
form of function $f(G)$. Finally, for specific theories which admit
the phantom phase where a final singularity may occur, the quantization
program should be developed (for recent related discussion, see
\cite{guido}).

\ack
%\subsection*{Acknowledgments}
We thank Jordi Miralda for reading this ms.


\section*{References}

\begin{thebibliography}{99}

\bibitem{nabhan}
Padmanabhan T {\it Preprint} astro-ph/0510492

\bibitem{mcinnes}
McInnes B 2002 {\it JHEP}\ {\bf 0208} 029 ({\it Preprint} hep-th/0112066)

\bibitem{phantom}
Faraoni V 2002 {\it Int.\ J.\ Mod.\ Phys.}\ D {\bf 64} 043514;
{\it Preprint} gr-qc/0404078; {\it Preprint} gr-qc/0506095;
Nojiri S and Odintsov S D 2003 {\it Phys.\ Lett.}\ {\bf B562} 147; ({\it Preprint} hep-th/0303117);
2003 {\it Phys.\ Lett.}\ {\bf BB571} 1 ({\it Preprint} hep-th/0306212);
Singh P, Sami M and Dadhich N 2003 {\it Phys.\ Rev.}\ D {\bf 68} 023522 ({\it Preprint} hep-th/0305110);
Gonzalez-Diaz P 2004 {\it Phys.\ Lett.}\ {\bf B586} 1 ({\it Preprint} astro-ph/0312579);
Sami M and Toporensky A 2004 {\it Mod.\ Phys.\ Lett.}\ {\bf A19} 1509 ({\it Preprint} gr-qc/0312009);
Gonzalez-Diaz P and Siguenza C 2004 {\it Nucl.\ Phys.}\ {\bf B697} 363 ({\it Preprint} astro-ph/0407421);
Chimento L P and Lazkoz R 2003 {\it Phys.\ Rev.\ Lett.}\ {\bf 91} 211301 ({\it Preprint} gr-qc/0307111);
2004 {\it Mod.\ Phys.\ Lett.}\ {\bf A19} 2479 ({\it Preprint} gr-qc/0405020);
Hao J and Li X 2005 {\it Phys.\ Lett.}\ B {\bf 606} 7 ({\it Preprint} astro-ph/0404154)

\bibitem{phantom2}
Nesseris S and Perivolaropoulos L 2004 {\it Phys.\ Rev.}\ D {\bf 70} 123529 ({\it Preprint} astro-ph/0410309);
Guo Z, Piao Y, Zhang X and Zhang Y 2005 {\it Phys.\ Lett.}\ B {\bf 608} 177 ({\it Preprint} astro-ph/0410654);
Anisimov A, Babichev E and Vikman A {\it Preprint} astro-ph/0504560;
Zhang X,  Li H, Piao Y and Zhang X {\it Preprint} astro-ph/ 0501652;
Wei Y 2005 {\it Mod.\ Phys.\ Lett.}\ A {\bf 20} 1147 ({\it Preprint} gr-qc/0410050);
{\it Preprint} gr-qc/0502077;
Dabrowski M and Stachowiak T {\it Preprint} hep-th/0411199;
Wu P and Yu H {\it Preprint} gr-qc/0509036

\bibitem{phantom3}
Wei H, Cai R.-G. and Zeng D {\it Preprint} hep-th/0501160;
Gumjudpai B, Naskar T, Sami M and Tsujikawa S 2005 {\it JCAP}\ {\bf 0506} 007
({\it Preprint} hep-th/0502191);
Andrianov A, Cannata F and Kamenshchik A {\it Preprint} gr-qc/0505087;
Wei H and Cai R.-G. {\it Preprint} astro-ph/0509328

\bibitem{QG}
Elizalde E, Nojiri S and Odintsov S D 2004 {\it Phys.\ Rev.}\ {\bf D70} 043539
({\it Preprint} hep-th/0405034).

\bibitem{quantum}
Nojiri S and Odintsov S D 2004 {\it Phys.\ Lett.}\ {\bf B595} 1 ({\it Preprint} hep-th/0405078);
Wu P-X and Yu H-W 2005 {\it Nucl.\ Phys.}\ {\bf B727} 355 ({\it Preprint} astro-ph/0407424);
Srivastava S {\it Preprint} hep-th/0411221

\bibitem{tsujikawa}
Nojiri S, Odintsov S D and Tsujikawa S 2005 {\it Phys.\ Rev.}\ D {\bf 71} 063004
({\it Preprint} hep-th/0501025)

\bibitem{stringy}
Nojiri S, Odintsov S D and Sasaki M {\it Phys.\ Rev.}\ D {\bf 71} 123509
({\it Preprint} hep-th/0504052);
Sami M, Toporensky A, Trejakov P and Tsujikawa S {\it Preprint} hep-th/0504154;
Calcagni G, Tsujikawa S and Sami M {\it Preprint} hep-th/0505193;
Amendola L, Charmousis C and Davis S {\it Preprint} hep-th/0506137;
Dadhich N {\it Preprint} hep-th/0509126;
Carter B and Neupane I {\it Preprint} hep-th/0510109

\bibitem{stringy1}
Arefeva I Ya, Koshelev A S and Vernov S Yu {\it Preprint} astro-ph/0412619;
{\it Preprint} astro-ph/0507067;
McInnes B {\it Preprint} hep-th/0502209

\bibitem{lambda}
Bauer F {\it Preprint} gr-qc/0501078;
Sola J and Stefancic H {\it Preprint} astro-ph/0505133;
Guberina B, Horvat R and Nikolic H {\it Preprint} astro-ph/0507666

\bibitem{EOS}
Nojiri S and Odintsov S D 2004 {\it Phys.\ Rev.}\ D {\bf 70} 103522 ({\it Preprint} hep-th/0408170);
Stefancic H 2005 {\it Phys.\ Rev.}\ D {\bf 71} 084024 ({\it Preprint} astro-ph/0411630);
{\it Preprint} astro-ph/0504518;
Szydlowski M, Godlowski W and Wojtak R {\it Preprint} astro-ph/0505202.

\bibitem{inh}
Nojiri S and Odintsov S D {\it Preprint} hep-th/0505215;
Capozziello S, Cardone V F, Elizalde E, Nojiri S and Odintsov S D {\it Preprint} astro-ph/0508350

\bibitem{GB}
Nojiri S and Odintsov S D {\it Preprint} hep-th/0508049

\bibitem{ANO}
Abdalla M C B, Nojiri S and Odintsov S D 2005 {\it Class.\ Quant.\ Grav.} {\bf 22} L35
({\it Preprint} hep-th/0409177)

\bibitem{modified}
Nojiri S and Odintsov S D 2003 {\it Phys.\ Rev.}\ D {\bf 68}, 123512 ({\it Preprint} hep-th/0307288)

\bibitem{DK}
Dolgov A D and Kawasaki M 2003 {\it Phys.\ Lett.}\ B {\bf 573} 1 ({\it Preprint} astro-ph/0307285);
Soussa M E and Woodard R P 2004 {\it Gen.\ Rel.\ Grav.} {\bf 36} 855 ({\it Preprint} astro-ph/0308114);
Nojiri S and Odintsov S D 2004 {\it Mod.\ Phys.\ Lett.}\ A {\bf 19} 627 ({\it Preprint} hep-th/0310045)

\bibitem{CNO}
Capozziello S, Nojiri S and Odintsov S D {\it Preprint} hep-th/0507182

\bibitem{multamaki}
Anarzguioui M, Elgaroy O, Mota D F and Multamaki T {\it Preprint} astro-ph/0510519

\bibitem{koivisto}
Koivisto T and Kurki-Suonio H {\it Preprint} astro-ph/0509422

\bibitem{guido}
Cognola G, Elizalde E, Nojiri S, Odintsov S D and Zerbini S
2005 {\it JCAP}\ {\bf 0502} 010 {\it Preprint} hep-th/0501096











\end{thebibliography}
\end{document}